\documentclass[12pt]{JHEP3}

\newif\ifPDFLATEX
\PDFLATEXfalse

\usepackage{amsmath,epsfig}
\usepackage{amssymb,amsfonts}
\usepackage{latexsym}
\usepackage{epsfig}

\relax
\renewcommand{\theequation}{\arabic{section}.\arabic{equation}}
\def\be{\begin{equation}}
\def\ee{\end{equation}}
\def\bea{\begin{eqnarray}}
\def\eea{\end{eqnarray}}
\def\hri#1#2{\href{http://arxiv.org/abs/#1}{[ArXiv:#1]#2}}
\def\hre#1#2{\href{http://arxiv.org/abs/#1/#2}{[ArXiv:#1/#2]}}

\newcommand\fverb{\setbox\pippobox=\hbox\bgroup\verb}
\newcommand\fverbdo{\egroup\medskip\noindent%
                        \fbox{\unhbox\pippobox}\ }
\newcommand\fverbit{\egroup\item[\fbox{\unhbox\pippobox}]}

\newcommand{\bear}{\begin{eqnarray}}

\newcommand{\eear}{\end{eqnarray}}
\newbox\pippobox

\def\6{\partial}

\def\a{\alpha}

\def\sp{\;\;\;,\;\;\;}

\def\sq
\def\a{\alpha}

\large
\title{$SU(5)$ orientifolds, Yukawa couplings, Stringy Instantons and Proton Decay}

\author{{\large Elias Kiritsis$^{1}$\footnote{On leave of absence from APC, Universit\'e Paris 7, (UMR du
CNRS 7164).
}, Michael Lennek$^{2}$ and Bert Schellekens$^{3,4,5}$ }
\\
~\\
$^1$Crete Center for Theoretical Physics\\
Department of Physics, University of Crete,
71003 Heraklion, Greece\\
~\\
$^2$CPHT, Ecole Polytechnique, CNRS,
 91128, Palaiseau, France\\
~\\
$^3$NIKHEF,
Science Park 105, 1098 XG Amsterdam,
The Netherlands\\
~\\
$^4$IMAPP, Radboud Universiteit Nijmegen, The Netherlands\\
~\\
$^5$Instituto de F\'\i sica Fundamental, CSIC, Madrid, Spain\\ }

\preprint{
CCTP-2009-15\\
CPHT-RR082.0809 \\
NIKHEF/2009-18\\}      

\abstract{We construct a large class of SU(5) orientifold vacua with tadpole cancellation both for the
standard and the flipped case.  We give a general analysis of superpotential couplings up to quartic order
in orientifold vacua  and identify the properties of needed Yukawa couplings as well as the baryon number
violating couplings.  We point out that successful generation of the perturbatively forbidden Yukawa
couplings entails a generically disastrous rate for proton decay from an associated quartic term in the
superpotential, generated from the same instanton effects. 
We search for the appropriate instanton effects that generate the missing Yukawa
couplings in the SU(5) vacua we constructed and find them in a small subset of them.}

\begin{document}

\section{Introduction}

String vacua involving open strings \cite{bs} have been seriously considered for the SM search
 after non-perturbative string dualities indicated that the heterotic string did not have the
  monopoly of interesting and
complex vacuum structure, \cite{wit}.  Interest in such vacua was enhanced by the observation that the
string scale  was less tightly constrained than in heterotic ones, \cite{wit2,lykken,anto}.

Orientifold vacua obtained a novel and important impetus after the realization \cite{akt,akt1,aiqu} that they
allowed a modular (bottom-up) approach in assembling the ingredients of the SM. This promoted local
constructions of D-brane stacks that could carry the SM spectrum and could be them embedded in full-
fledged string compactifications.  There are many distinct ways of embedding the Standard Model group
into that of quiver gauge theories, which appear in the context of orientifolds and these are reviewed in
\cite{d-review-1}-\cite{d-review-5}.

The prototype of the modular construction approach was implemented via RCFT techniques.  The
orientifolds were constructed from Gepner models (studied earlier in \cite{an1}-\cite{an5}), using the
algorithmic techniques of RCFT developed in \cite{Fuchs:2000cm}.  In the first search in \cite{dhs}, vacua
realizing the Madrid incarnation \cite{imr} of the Standard Model were analyzed.  They provided the
largest collection of vacua (tadpole solutions) to date,  chirally realizing the (supersymmetric) SM.

In the same context, a more general search was done where all possible embeddings of the SM in four
stack configurations was analyzed \cite{adks}.
A total of 19345 chirally distinct top-down spectra were found,
 that comprise so far the
most extensive such list known in string theory \cite{adks}.
 For 1900 of these 19345 spectra at least one tadpole solution was also found (no further attempts at
solving tadpoles were made
 once a solution was found for a given chirally distinct type).
 The wealth of tadpole solutions can only be compared to a recent extensive list from the $Z_6'$
orientifold, \cite{gmeiner}, although
 even that set appears to cover far fewer distinct possibilities.
 Not all regions of moduli space are rich in SM-like vacua though. The $Z_2\times Z_2$ orientifolds
\cite{billion}
 and the free-fermionic orientifolds \cite{free} although they contain a large number of vacua, seem to be
SM-free.
Recently tachyon-free tadpole-free non-supersymmetric vacua have been searched for \cite{schellekens-ns} in
Gepner models.
No solutions were found, although there do exist many tachyon-free
non-supersymmetric ``local" configurations with
uncancelled tadpoles.

On a different note, heterotic and orientifold vacua look generically different at least in one direction.
Although in heterotic vacua, there is a generic
underlying GUT structure, this is generically not the case in orientifold vacua.
In a generic orientifold construction the SU(3) and SU(2) groups originate in generically distinct D-brane
stacks, without  an  {\it a priori}
relation of their respective gauge couplings at the string scale. Moreover, as was first analyzed
in detail in \cite{akt}, the hypercharge gauge group is always a linear combination of U(1)'s from different
brane stacks, although this linear combination
may vary from orientifold to orientifold. The general such hypercharge embedding was
classified in \cite{adks} in terms of a real number $x$ that is typically discrete.

This characteristic structure is responsible for some unique generic properties of SM-like orientifold vacua
in particular  the presence of at least one anomalous U(1)
gauge boson in the standard model stack\footnote{The generic number is three, \cite{akt,akt1,adks}.},
which mixes because of electroweak
symmetry breaking with the photons and $Z_0$. This mixing can be substantial and observable if the
associated anomalous gauge boson is
light, \cite{akt,g-2,giiq,cik,abdk}. This can happen both when the string scale is at the TeV scale, as well as
when it is higher,
 if there are large cycles in the compactification manifold, \cite{au1}.
 Another property relates to the novelty and richness of patterns and mechanisms for the generation of the
hierarchy of masses
 \cite{ib,KST,leontaris,akl,myr},
  a fact that   is welcome in the search for realistic vacua in string theory.

It is admittedly true that there are several indications favoring embedding the SM in a unified gauge
group.
 They include the apparent unification of coupling constants at a ``GUT" scale $M_{GUT}\sim 10^{16}$
GeV as well as
 the appearance of the same scale in seemingly unrelated sectors (neutrino masses, dark matter etc).
A direct attempt to generate the popular GUT groups\rlap,\footnote{Pati-Salam groups are natural in
orientifolds and many such vacua have been found.
The earliest examples are in \cite{ps1,ps2,ps3} while  the largest list of tadpole solutions without chiral
exotics
 is in \cite{adks}.}
 (SU(5) and its relative ``flipped SU(5)", SO(10), and E$_6$) in orientifolds indicates that only SU(5) is
possible. Both SO(10) and E$_6$ contain spinor-like representations and these cannot be constructed
perturbatively in the context of orientifolds.
They can be constructed non-perturbatively however, as the duality with the heterotic string \cite{wit} and
direct constructions have indicated
\cite{sen}. Although this can be done,  the models lose their weak-coupling appeal as the
non-perturbative states are very heavy in perturbation theory.

 On the other hand,  SU(5) and related constructions can be achieved by embedding the SU(5) in a U(5)
stack of branes.
 There are even other unified constructions that have not been discussed in the GUT literature but appear
naturally in orientifolds.
 An interesting example is a U(6) super-unified tadpole solution  found in \cite{adks} that through three
distinct symmetry breakings
  can produce an SU(5), flipped SU(5)
 or Pati-Salam intermediate group and spectrum.
 An early supersymmetric SU(5) example was described in \cite{cv1} and several others were constructed
in \cite{susy}.
 They all contained chiral exotics.
Other non-supersymmetric SU(5) examples were found but as usual in these cases tadpole conditions
were not satisfied \cite{non-susy1}, \cite{non-susy2}.
The first supersymmetric examples without chiral exotics and satisfying all tadpole conditions were
constructed in \cite{adks}.

Although SU(5) orientifold vacua have a simple structure and produce easily the appropriate spectrum
including right-handed neutrinos
they suffer from an important ailment.
 It was fits pointed out in \cite{non-susy1} (see also \cite{bere} and \cite{adks}), that the top Yukawa coupling is absent in
perturbation theory as it carries a non-zero charge under $U(1)_5$ (the U(1) factor of the U(5) group).

There are two possible ways to generate such a Yukawa term. Both of them break in a way the $U(1)_5$
global symmetry.
A first possibility is turning-on fluxes that break $U(1)_5$. Such a mechanism has not so far been
explored in detail
due to the difficulty of constructing realistic vacua with non-trivial fluxes.
The second possibility relies on the fact that generically
the $U(1)_5$ gauge symmetry is of the anomalous type and the associated global
 symmetry is expected to be broken by non-perturbative (instanton) effects.
 This is the route we will pursue here.

Non-perturbative instanton effects in string theory have been discussed early on,
while the non-perturbative dualities were explored, \cite{ov}.
In particular, D-instanton effects in open string theory could be mapped to perturbative string effects on a
dual heterotic side,
\cite{bfkov}, and this gave  the first glimpse
into the structure of the D-instanton corrections (for an early review see \cite{i-review}).
Several years later, the structure of gauge instantons were elucidated using D-brane techniques,
\cite{stringinst-1}.
 Lately, D-instantons have been argued to provide non-trivial contributions to couplings protected
otherwise by anomalous U(1) symmetries, like
 neutrino masses, \cite{stringinst-2} motivating a resurgence of interest whose output has been reviewed
in \cite{instanton-review}.
The first global example analyzed, that provided non-zero instanton contributions, was based on the
$Z_3$ orientifold, \cite{bk}.
It provided the generation of an ADS superpotential, mass terms for chiral multiplets, that together lead to
supersymmetry breaking contributions if
the closed string moduli are stabilized,  \cite{susybr}.
Although at a general point in moduli space the gauge group is SU(4), there are enhanced regions where
the group is SU(5)  with a spectrum
of three antisymmetric
chiral multiplets, 3 $\bar 5$s as well as 3 pairs of ($5+\bar 5$) Higgses, \cite{iu}.
In this phase, the same instanton generates the top-like Yukawa
couplings. Upon further Higgsing to SU(4) these match the instanton generated mass terms computed in \cite{bk}.
Several further works analyzed the structure of instanton corrections further \cite{cam}-\cite{ange} 
and in particular the generation of the top Yukawa couplings in SU(5) orientifolds both at
the local and global level
\cite{refs,Blum}.

In the context of RCFT orientifold constructions of Madrid-like SM embeddings,  \cite{imr}, a search for
instanton effects was
done, in order to track neutrino mass generation. The experience from such a search is that  RCFT vacua,
having typically enhanced symmetries,
possess instantons with typically large number of zero modes. Therefore instanton contributions to the
superpotential are atypical, and indeed no
single instanton contribution  was found in \cite{isu}.

The results of the paper can be summarized as follows:

\begin{itemize}

\item We analyze orientifold vacua with SU(5) gauge group, realizing SU(5) or flipped SU(5) grand
unification.
We construct many tadpole solutions from Gepner model building blocks using the algorithm
developed in \cite{adks}.
We found all such top-down constructions as well as tadpole-free vacua, with one extra observable brane
of the U(1) or O(1) type.
This is one small subset (but the simplest) of the SU(5) configurations found in \cite{adks}.

\item We give a general analysis of possible terms in the superpotential of such vacua, up to quartic order, and
classify them according to their
fatality ( baryon and lepton violating interactions which are relevant or marginal), and  usefulness
(Yukawa coupling).
We have classified which terms can or must be generated by instanton effects.
As is well known, the top Yukawa's in SU(5) and the bottom in flipped SU(5) must be generated from
instantons (in the absence of fluxes).

\item We find that in flipped SU(5) vacua, B-L cannot be anomalous as it participates in
the hypercharge.
This forbids all dangerous terms, but it is necessarily broken when the SU(5) gets broken at the GUT scale.
The proton decay generated is estimated to be typically small.

\item In U(5)$\times$U(1)  vacua, instanton effects must generate the top Yukawa couplings, and at the same
time they break the B-L symmetry.
Successful vacua, have either a $Z_2$ remnant of the B-L symmetry acting as as R-parity and forbidding
the dangerous terms, or
such terms may have exponentially suppressed instanton contributions. In the second case they are viable
if the exponential factors are
sufficiently suppressed.  We provide several tadpole solutions of the first
case where instantons generate the top Yukawa's, but preserve a $Z_2$ R-symmetry.

\item  U(5)$\times O(1)$ vacua are problematic on several grounds and need
extra symmetries beyond those that are automatic, in order to have a chance of not
being outright excluded. This is related to the absence of natural R-symmetries or gauge symmetries that
 will forbid the dangerous low-dimension baryon-violating interactions.

\item A generic feature of all SU(5) vacua is that the same instanton the generates the non-perturbative quark
Yukawa coupling also generates
the $10~10~10~\bar 5$ in the superpotential. This is a second source of proton decay, beyond the
classic one emanating from the
Higgs triplet times the appropriate Yukawa coupling. Generically, the size of this contribution to proton
decay
is $10^5 ~{M_T\over M_s}$ larger than the conventional source in flipped SU(5) model, ($M_T$ is the
triplet Higgs mass).
This  signals severe phenomenological trouble
and calls for important fine tuning. In the SU(5) case the size is 30 times smaller, but that does not evade
the need for fine tuning.

\item  We have searched for appropriate instantons that would generate the perturbatively forbidden quark
 Yukawa couplings in the SU(5) vacua we have constructed. We found the appropriate instantons with the
correct number of zero modes in 6 relatives
 of the spectrum Nr. 2753. We have also searched for all other instantons that could generate the bad
terms in the superpotential and found none.
 This translates into the existence of a $Z_2$ R-symmetry that protects from low-dimension baryon and
lepton-violating couplings.

\end{itemize}

The structure of this paper is as follows:

In section 2 we describe the search for SU(5) vacua with tadpole cancellation with at most one extra
observable stack using the RCFT
of Gepner models.

In section 3 we give a general analysis of superpotential terms up to quartic order in such vacua, their
relevance for baryon and lepton number violation as well as
the possibility of their  generation via non-perturbative effects.

In section 4 we give an analysis of the relevant instanton zero modes and their impact in the generation of
terms in the superpotential.

In section 5 we provide an analysis of the relative effects in proton decay of two superpotential operators
whose generation by instantons is correlated

 In section 6, we search the RCFT vacua found for the instantons appropriate to generate the missing
quark Yukawa couplings.

 Section 7 contains our conclusions.

 In Appendix A we provide the complete spectrum of the class of tadpole solutions with the requisite
instanton effects.

\section{Explicit Constructions}
\label{Gep}

In Ref.~\cite{adks}, a methodology for identifying self-consistent semi-realistic string models was
developed.  This methodology
was employed on a set of string vacua constructed using RCFT techniques on Gepner models and a host
of examples were
presented.  In this paper, we are primarily interested in orientifold vacua with an SU(5) GUT group (both
standard and flipped).
We are also interested on the possibility of generating the appropriate Yukawa couplings
(forbidden perturbatively) by string instantons. Moreover we will also analyze some issues
related to this mechanism and in particular the issue of proton decay.
For this we will revisit some of the models originally presented in \cite{adks}.

The methodology developed in \cite{adks}, starts with a variation the bottom-up approach developed in
\cite{akt,aiqu}. Instead of
geometric brane configurations, RCFT boundary state combinations are searched for that give rise to a
spectrum of interest (usually the MSSM
or a unified extension of the latter). Then an attempt is made to find additional boundary states that
provide a ``hidden sector" that can
cancel the tadpoles. This method was pioneered in \cite{dhs} and is based on the boundary state
formalism presented in \cite{fhssw}, which in
its turn is based on earlier work, such as \cite{Pradisi:1996yd,Pradisi:1995pp} and \cite{Cardy:1989ir}.
This method provide bona-fide string vacua that have low-energy limits
consistent with the MSSM.  We shall briefly summarize the relevant points for the subset of these string
models considered in the present work:
\begin{itemize}
\item{The visible sector is required to consist of three or fewer stacks of branes where the $SU(5)$ arises
from exactly one stack.}
\item{The chiral spectrum for the visible sector should reduce to three generations of the
MSSM, once the gauge group is reduced to $SU(3)\times SU(2) \times U(1)$.}
\item{No chiral exotics are present in the spectrum.  Chiral exotics are defined as {\it any} states that are
chiral
with respect to the standard model gauge group and that do not fit in the usual three families.}
\end{itemize}
Here ``visible sector" is defined as the set of branes that contribute to the standard model gauge group
and/or the
charged quarks and leptons (some right-handed neutrinos may also originate from the visible sector, but
are not required).

Note that the set of chiral states in the visible sector may be larger than just the three standard model
families. We are
only requiring that the superfluous ones become non-chiral under a group-theoretical reduction to $SU(3)
\times SU(2) \times U(1)$.
For an example, see the spectrum presented in table 5 below. However, the other examples considered
in the present paper, and
the vast majority of Madrid type models considered in \cite{dhs} have no superfluous chiral states
whatsoever.

 In both \cite{dhs} and \cite{adks} chiral states that are charged under both the visible and the hidden
sector
were forbidden, even if they reduce to non-chiral standard model states. This is a bit more restrictive than
the conditions
imposed on the visible sector. For example, the combination $(5,r_1)+(\bar 5,r_2)$, where $r_1$ and
$r_2$ are distinct
hidden sector representations of equal dimension, would not be allowed in both papers, even though it
reduces to a vector-like $SU(5)$
representation.  Obviously a mere group-theoretic reduction is not sufficient to give a mass to these
vector-like state. One would
have to get into the details of hidden sector dynamics in each individual case to see if such a model is
viable, and for this reason
lifting this requirement is unattractive. However, in order to be as complete as possible we have lifted this
requirement in the present paper.

Explicit examples of models are presented below.

\subsection{The $SU(5)$ orientifolds\label{mod}}

In a previous study focusing on Gepner models~\cite{adks}, $SU(5)$ models satisfying the criteria listed
above
were presented.  In the full set of 1900 chirally distinct tadpole solutions there are
494 cases where the $SU(3)$ and $SU(2)_{W}$ branes are identical. This implies an extension of the
standard model
group to at least $SU(5)$ (in some cases $SU(5)$ is a subgroup of a larger unitary group). Two spectra
are called
``chirally distinct" if the visible sector gauge groups are different, if the matter that is chiral with respect to
the gauge group
is different, or if different $U(1)$ bosons acquire a mass through axion mixing. Consequently, an $SU(5)$
model is regarded
as distinct from the model obtained by splitting  the $U(5)$ stack into a $U(3)$ and a $U(2)$ stack. In
some cases, both possibilities
exist.

The simplest orientifold realizations of $SU(5)$ models consist of one $U(5)$ brane stack, plus one
additional brane, with Chan-Paton multiplicity 1, intersecting that stack. Matter in the $(10)$ of $SU(5)$
arises from chiral anti-symmetric
tensors, where matter in the $(\bar 5)$ comes from intersections of the $U(5)$ brane and the additional
brane.
The search criteria of \cite{adks} allow for even more general $SU(5)$ models. Instead of getting three copies of
the
$(\bar 5)$ from a triple intersection, one may obtain each family from a separate brane intersecting the
$U(5)$ stack,
and/or get some of the family multiplicity by allowing higher CP-multiplicities. There is indeed a large
variety of such
more complicated spectra in the database of \cite{adks}, but we consider here only the ones with a single
additional brane.

The additional brane can be either $O(1)$ or $U(1)$.
In the following table we list the distinct models and how often they occurred. Note that this refers to brane
configurations prior to attempting to cancel tadpoles (named top-down models in \cite{adks}).
We note also that counting the number of distinct
models has some subtleties.  It is likely that some of them are
merely distinct points in the same moduli space. Moving around in the moduli space would then
provide the differences in vector-like matter.
\TABLE{
\begin{tabular}{|c||c|c|c|}
\hline
\hline
{Nr.}& Frequency & CP group  & Gauge group\\
\hline
617 & 16845 & $U(5) \times O(1)$ & $SU(5)$\\
2753 & 1136 & $U(5) \times U(1)$ & $SU(5) $\\
2880 & 1049 & $U(5) \times U(1)$ & $SU(5) \times U(1)$\\
6580 & 146 & $U(5) \times U(1)$ & $SU(5)$\\
14861 & 12 & $U(5) \times U(1)$ & $SU(5)$\\
\hline
\end{tabular}
\label{ADKSspectra}
\caption{List of the $SU(5)$ models with a single additional brane.
The second column show the
number of such spectra (modulo non-chiral matter) that was found. The first column is the number used
to refer to these spectra, and is equal to the position of the spectrum on the full list of \cite{adks}, sorted by
frequency.
}
}

The unitary phase of the $U(5)$ stack is anomalous, and hence the corresponding gauge boson always
acquires a mass. However, in some of the $U(5) \times U(1)$ models a linear combination of the
$U(5)$ and $U(1)$ phases is anomaly-free, and
may or may not acquire a mass.  It remains massless in model type
2880, whereas it is anomaly-free, but not massless in model type 2753.
The existence of this anomaly free combination
corresponds to the possibility to interpret the model as a flipped $SU(5)$ model. Hence model type 2880
has two distinct interpretations, either as a normal or as a flipped $SU(5)$ model. In both interpretations
there
is necessarily an additional massless $U(1)$ gauge boson in the exact string string spectrum,
corresponding to
$B-L$. In the other four model types the gauge group is exactly $SU(5)$, plus a hidden sector that may be
required for
tadpole cancellation.

Tadpole canceling hidden sectors were found for model types 617, 2753 and 2880. In \cite{adks} these
solutions
were not optimized for simplicity; for each model type just one tadpole solution was collected.

We have done a systematic
analysis of all 16845 $U(5)\times O(1)$ models, and the results are as follows. First we tried to solve the
tadpole
conditions allowing chiral matter between the observable and hidden sector. For 15499 of these 16845
we were
able to show that no such solution exists, for 641 we did find a solution, and 705 cases were inconclusive:
the
tadpole cancellation equations were too complicated to decide if there is a solution. The algorithm used
for solving
the tadpole conditions consists of two parts: an exhaustive search for all solutions with up to four hidden
branes, plus a different algorithm allowing in principle an arbitrary number of hidden branes. The latter
search
was limited in time. If it is terminated prematurely such a case is labeled ``inconclusive".

Next we tried to solve the tadpole conditions under the more restrictive condition that only non-chiral
observable-hidden
matter is allowed (the same condition as used in \cite{adks} and \cite{dhs}). Of the 641 cases that had
solutions in the
previous search, 521 had no non-chiral solutions, 109 did have non-chiral solutions, and 11 were
inconclusive.  Of the
705 previously inconclusive cases, 508 had no non-chiral solutions, 64 had non-chiral solutions, and 133
were still
inconclusive. These numbers give some idea about the success rates of of attempts to solve the tadpole
conditions.

The simplest solution found for model type 617 is shown in table \ref{ADKSmodOne}.

\TABLE{
\begin{tabular}{|c||c|c|c|}
\hline
\hline
& \multicolumn{2}{c|}{Visible Sector}& Hidden Sector\\
\hline
Multiplicity & $U(5)$ & $O(1)$ & $O(1)$\\
\hline
3(3+0) & A & 0 & 0\\
5(4+1) & V$^*$ & V$^*$ & 0 \\
8(4+4) & V & 0 & V \\
\hline
2(1+1) & V & V & 0\\
8(4+4) & S & 0 & 0 \\
3 & Adj & 0 & 0 \\
\hline
1 & 0 & A & 0  \\
3 & 0 & V & V  \\
2 & 0 & S & 0  \\
4 & 0 & 0 & S  \\
4 & 0 & 0 & A  \\
\hline
\hline
\end{tabular}
\label{ADKSmodOne}
\caption{The particle spectrum of the simplest $SU(5)$ model found.  The visible sector consists of $U
(5)$ and $O(1)$ with a hidden sector consisting only of $O(1)$.  The transformation properties under the
various groups are listed on the table where a ``V" refers to a vector, ``S" is a symmetric tensor, ``A" is an
antisymmetric tensor, ``Adj" is an adjoint, and $0$ is a singlet. In the first column ``M+N" means M copies
of
the representation, plus N copies of its complex conjugate. }
}

For model type 2880 the results are as follows.
The 1049 configurations split into two classes: one where the dilaton tadpole condition is already
saturated, so that there is
no room for a hidden sector, and one where a hidden sector is required. In the former case one can only
hope
that the remaining tadpole conditions are solved as well. This class contains 437 configurations, and {\it
all} of them
turn out to satisfy all tadpole conditions! All of them are closely related, and occur for the same MIPF of
tensor product (1,4,4,4,4).
Only a few of these 437 spectra are distinct, and the simplest one is shown in table \ref{ADKSmodTwo}.
Note that this is
a different spectrum then the one presented in \cite{adks}, because in that paper no attempt was made to
look for the
simplest version of a spectrum.
The column ``Variations" list the other values for the total multiplicity that were found (though not
uncorrelated).

The other 612 configurations require a hidden sector, and in only 10 of them it can indeed be found. All of
these have
chiral hidden-observable matter, and therefore they do not satisfy the original requirements of \cite{adks}.
Indeed, in that
paper no such solution was found.


\TABLE{\hbox{~~~~~~~~~~~~~~~~~~~~}
\begin{tabular}{|c||l||c|c|}
\hline
\hline
& & \multicolumn{2}{c|}{Visible Sector }\\
\hline
Multiplicity & Variations & $U(5)$ & $U(1)$ \\
\hline
3(3+0) & 3,7 & A & 0 \\
3(3+0) & 3,5 & V$^*$ & V$^*$  \\
3(3+0) & 3,7,11 &0 & S \\ \hline
4(2+2) & 4,8,12 & V & V$*$ \\
8(4+4) & 4,8 & S & 0 \\
3 & 3,5,7,9 & Adj & 0  \\
3 & 3,5 & 0 & Adj  \\
\hline
\hline
\end{tabular}\hbox{~~~~~~~~~~~~~~~~~~~~}
\label{ADKSmodTwo}
\caption{The particle spectrum of the  $SU(5) \times U(1)$ model nr. 2880. This satisfies all tadpole
conditions without a hidden sector.}
}

For model nr. 2753 only a rather complicated solution with seven hidden sector factors was found in
\cite{adks}. This solution
was just a sample, the first one that was encountered.
We have now  scanned all
1136 models of this type, and nothing simpler  was found. Of the 1136 models, six turned out to admit a
solution, and the
other 1130 did not. All these solutions are similar to the first sample found.
They all have a hidden sector $U(5)\times Sp(4) \times U(2) \times O(2)^2 \times U(1)^2$,
a large number of non-chiral exotics, and some chiral fermions entirely within the hidden sector. Further
details
will not be presented here. Instead, we will present a spectrum for this model
with instanton branes in section (\ref{InstantonSearch}).

For the remaining two model types no tadpole solution was found.  Below we display the
``local" spectra of these models, {\it i.e.} the standard model configuration without
tadpole cancellation. As always, this is just the first sample found, without any attempt
to optimize or simplify the non-chiral spectrum. The spectrum for model nr. 6580 is remarkably simple and
shown in table
\ref{ADKSmodFour} .

\TABLE{\hbox{~~~~~~~~~~~~~~~~~~~~}
\begin{tabular}{|c||c|c|}
\hline
\hline
& \multicolumn{2}{c|}{Visible Sector  }\\
\hline
Multiplicity & $U(5)$ & $U(1)$ \\
\hline
3(3+0) & A & 0 \\
2(2+0) & V$^*$ & V$^*$  \\
1(1+0) & V$^*$ & V \\
1(1+0) & 0 & S \\
\hline
\hline
\end{tabular}\hbox{~~~~~~~~~~~~~~~~~~~~}
\label{ADKSmodFour}
\caption{The local spectrum of the  $SU(5) \times U(1)$ model nr. 6580. No global version of this model
was found.}
}

Note the complete absence of any non-chiral matter, an extremely rare feature. However, this
also implies the absence of any Higgs candidates  for breaking $SU(5)$ or for breaking
$SU(2) \times U(1)$. The last model type, Nr. 14861, of which only 12 examples were seen in \cite{adks},
is shown in table \ref{ADKSmodFive}. Note that
this spectrum contains 6 $(\bar 5)$'s and 3 $(5)$'s, which are chiral with respect to the additional $U(1)$.
However, since this $U(1)$ is
not part of the standard model gauge group, it {\it does} satisfy our criteria.
\TABLE{\hbox{~~~~~~~~~~~~~~~~~~~~}
\begin{tabular}{|c||c|c|}
\hline
\hline
& \multicolumn{2}{c|}{Visible Sector }\\
\hline
Multiplicity & $U(5)$ & $U(1)$ \\
\hline
3(3+0) & A & 0 \\
6(6+0) & V$^*$ & V$^*$  \\
5(4+1) & V & V$^*$ \\
15(12+3) & 0 & S \\
1 & Adj & 0 \\
4 & 0 & Adj \\
\hline
\hline
\end{tabular}\hbox{~~~~~~~~~~~~~~~~~~~~}
\label{ADKSmodFive}
\caption{The local spectrum of the  $SU(5) \times U(1)$ model nr. 14861. No global version of this model
was found.}
}

\section{Yukawa terms and other relevant couplings\label{yuk}}

\TABLE{
\begin{tabular}{|c|c|c||c|c|}
\hline
\hline
$SU(5)~ {\rm Rep.}$ & $U(1)_1$ & $U(1)_5$& Flipped $SU(5)$ Matter Content& $SU(5)$ \\
\hline
10 & $0$ & +2& $(Q, d_L^c, \nu_L^c)$& $(Q, u_L^c, e_L^c)$ \\
$\overline{5}$ & $-1$ & $-1$& $(L, u_L^c)$& $(L, d_L^c)$\\
1 & +2 & $0$& $(e_L^c)$& $(\nu_L^c)$\\
$5_H$ & $-1$& $+1$& $(H_d, T_d)$& $(H_u, T_u)$\\
$\overline{5_H}$ & $+1$& $-1$&$(H_u, T_u)$&$(H_d, T_d)$\\
\hline
\hline
\end{tabular}
\caption{The manner of embedding one generation of the SM into $SU(5)$ multiplets with their
respective $U(1)_{1,5}$ charges for the typical string realization.}
\label{Matter}}

There are several couplings in the superpotential whose size is crucial  for
acceptable low-energy physics.\footnote{Several related issues about some of these terms have been discussed in \cite{Blum}.}
We will list them below up to quartic order.
\begin{itemize}

\item The $\bar 5 5_H$ term gives rise to relevant lepton number violating interactions that can kill models
instantly.
 In model building it is typically forbidden by an R symmetry.

\item The $1~\bar 5 5_H$ term is a Yukawa coupling. The $1$ is the singlet that plays
 the role of the right-handed neutrino
in SU(5) models and the lepton singlet in flipped SU(5) models.
In both cases this term should be present to generate the appropriate mass.

\item The $10\bar 5\bar 5$ term,  where the $\bar 5$'s are matter,
generates dimension four operators that break baryon and lepton number and are instantly fatal
unless the coupling is exponentially suppressed.
These terms are usually forbidden by advocating an R-symmetry.

\item The $10\bar 5\bar 5_H$ is a standard Yukawa coupling that must appear
 with appropriate coefficients, as it generates the masses of half of the quarks and leptons
 (which ones depends whether we are in SU(5) or flipped SU(5).)

\item The $10\bar 5_H\bar 5_H$ is generating couplings between the light Higgs and the singlet.
For a single Higgs this is zero by symmetry. It contributes in the presence of more than one pairs of
Higgses.

\item The $1010 5_H$ term is a standard Yukawa coupling that must appear
 with appropriate coefficients, as it generates the masses of the other half of the quarks and leptons
  (which ones again depends whether we are in SU(5) or flipped SU(5).)

\item $101010\bar 5$ is a term that can be important for proton decay.

\item There is also an associated term $101010\bar 5_H$. This term seems relatively innocuous as we
will describe below.

\end{itemize}

We will now discuss the status of all of these terms in the various realizations of SU(5) orientifold models.

\subsection{Flipped SU(5) from $U(5)\times U(1)$ orientifolds\label{flsu}}

The standard generic spectrum in an $U(5)\times U(1)$ orientifold is given in table \ref{Matter}.
We may write the important U(1) symmetries in this context using the standard diagonal SU(5) generator
\be
W=\left(\begin{matrix} -{1\over 3} &0&0&0&0\cr
0&-{1\over 3}&0&0&0\cr
0&0&-{1 \over 3}&0&0\cr
0&0&0&{1\over 2}&0\cr
0&0&0&0&{1\over 2}\cr\end{matrix}\right)
\ee
as
\be
Y={Q_X-W\over 5}\sp Q_{B-L}={Q_X+4W\over 5}\sp Q_B={Q_5-2W\over 5}\sp Q_X={5Q_1+Q_5\over 2}
\ee
Above, $Q_5$ is the U(1) charge  of the U(5) stack, and $Q_1$ is the charge of the U(1) stack. Both are
normalized to integers.

To assess further the possibility of using instantons to generate the missing Yukawa's we will keep track of
the $U(1)_X$ charges.
The reason is that the potential instanton effects violating $U(1)_X$ are severely constrained by the fact
that
$U(1)_X$ participates in the Standard Model hypercharge $Y$.

The other generator participating in $Y$ is the SU(5) generator $W$ that is traceless,
and cannot therefore become massive nor can be violated by instanton
effects . As the same should be true for Y, (otherwise this is not acceptable in the SM)
 we conclude that for a string vacuum to be a viable SM candidate,
the $Q_X$ U(1) symmetry must be massless and therefore it should not be violated by instanton
effects\footnote{Masses for U(1) symmetries in string theory appear via the mixing
with various closed string forms. The appearance of a mass breaks the U(1) gauge symmetry but
 not the global U(1) symmetry. However we do not expect exact global (internal)
symmetries in string theory.
Indeed,   it is the defects charged under the same forms that appear as instanton effects
 violating the global U(1) symmetry, breaking it to a discrete subgroup.}

\begin{itemize}

\item The $\bar 5 5_H$ term has charge $(-2,0,-5)$ under ($U(1)_1$,$U(1)_5$, $U(1)_X$) and is
therefore forbidden.
As it is charged under $U(1)_X$. It might be generated only if $U(1)_X$ breaks spontaneously.

\item The $1~\bar 5 5_H$ term is a standard Yukawa coupling that gives mass to the lepton singlet
and it is perturbatively allowed as it is uncharged under ($U(1)_1$,$U(1)_5$).

\item The $10\bar 5\bar 5$ term has charge $(-2,0,-5)$ under ($U(1)_1$,$U(1)_5$, $U(1)_X$) and is
therefore forbidden.

\item The $10\bar 5\bar 5_H$ term is a standard Yukawa coupling that gives mass to the top quark
and is perturbatively allowed as it is uncharged under ($U(1)_1$,$U(1)_5$).

\item The $10\bar 5_H\bar 5_H$  has charge $(2,0,5)$ under ($U(1)_1$,$U(1)_5$,  $U(1)_X$) and is
therefore forbidden.

\item The $1010 5_H$ term gives masses to the bottom quark and the right-handed neutrino. It has
charges $(-1,+5,0)$ under ($U(1)_1$,$U(1)_5$, $U(1)_X$).
It is perturbatively forbidden and can only be generated by instantons.

\item The $101010\bar 5$ term has charges $(-1,+5,0)$ under ($U(1)_1$,$U(1)_5$, $U(1)_X$), and is
therefore perturbatively forbidden.
The charge structure is however the same as the previous Yukawa and later on we will argue
that such a term is generated by the same instanton effects.

\item The $101010\bar 5_H$ term has charges $(+1,+5,5)$ under ($U(1)_1$,$U(1)_5$, $U(1)_X$) and is
forbidden by the $U(1)_X$ symmetry.

\end{itemize}

Of all the couplings that are perturbatively forbidden,
$\bar 5 5_H$, $10\bar 5\bar 5$, $10\bar 5_H\bar 5_H$, $101010\bar 5_H$, $1010 5_H$, $101010\bar
5$,
the first four have $Q_X=\pm 5$ while the two last ones have $Q_X=0$.
This is not a surprise as $Q_X$ is intimately related to B-L and therefore forbids  the first four terms.

The upshot of the previous discussion is, that in flipped SU(5) orientifold vacua, if the hypercharge is
massless, there is necessarily a massless $U(1)_{B-L}$ gauge boson associated to the gauged B-L
symmetry. This can also be rephrased in the opposite way:
in order for a $U(5)\times U(1)$ configuration to reproduce the SM, the
$U(1)_{B-L}$ must be unbroken at high energy.
At some energy, $U(1)_X$ and B-L must be broken by a vev.
This is indeed what happens at the GUT scale as in flipped SU(5) models the
appropriate GUT symmetry breaking
happens when a Higgs multiplet that transforms as the 10 obtains a vev to
break SU(5)$\times $U(1) to SU(3)$\times
$SU(2)$\times $U(1).
At the same time it breaks $U(1)_X$.
We will denote by $10_H$ and $\underline{10}_H$ such Higgs fields.

Let us consider the potential generation of the most relevant ``unwanted term" in the superpotential
namely $\bar 5 5_H$ with charges (-2,0,-5)
under $(U(1)_1,U(1)_5,U(1)_X)$. As $U(1)_X$ is not broken by instantons the leading term in the
superpotential that can generate
$\bar 5 5_H$ after symmetry breaking is $\bar 5 5_H(10_H)^5$ with charges (-2,10,0).
This has now $Q_X$ charge zero and is therefore allowed, but is perturbatively forbidden by the
anomalous U(1) symmetry that forbids also the bottom
Yukawa coupling $10105_H$ with charges (-1,5,0). Therefore if there is an instanton that generates
the relevant Yukawa coupling $h_{10105_H}\sim e^{-S_{inst}}$, it is also plausible
(although it is not guaranteed\footnote{In the brane picture of instantons this
would amount to a O(2) or Sp(2) bound state. Although this will have too many
zero modes to contribute to the
superpotential there could be points in its moduli space where the symmetry is
broken and the associated zero modes are lifted along the lines of
 \cite{ura}.}) that there is also an instanton with charge violation
(-2,10,0)
that will generate the $\bar 5 ~5_H~(10_H)^5$ term.

Assuming this worst scenario case, we can therefore estimate its strength as
 $h_{10105_H}^2{<10>^5\over M_s^4}$, assuming that the instanton contribution
is the square of an (-1,5,0) instanton. This gives an effective scale
\be
\mu_{\bar 55_H}\sim  h^2_{10105_H}\left({M_{GUT}\over M_s}\right)^{4}~M_{GUT}~~\sim ~~10^{-4}~~
{\rm GeV}
\ee
To obtain the estimated value we have taken the Yukawa coupling of the strange quark as the central
value
for $h_{10105_H}\sim h_s\sim 4\times 10^{-4}$, ${M_{GUT}\over M_s}\sim 10^{-3}$ and $M_{GUT}\sim
10^{15}$ GeV.
This term in the IR generates an $L\bar H$ term that can be rotated into the lepton number violating marginal
interactions
by rotating in the space of lepton doublets and Higgs. If we assume a $\mu$ term of order the EW scale
then the angle of rotation
is  $\theta\sim {\mu_{\bar 55_H}\over \mu}\sim 10^{-6}$ and therefore well below the limits on such
couplings from lepton violation
\cite{DH,D}.

Similar arguments apply to the baryon violating $10\bar 5\bar 5$ term that descends from the $10\bar
5\bar 5 (10_H)^5$ term. The induced
dimensionless effective coupling
of the $10\bar 5\bar 5$ term in the superpotential is then
\be
h^{eff}_{10\bar 5\bar 5}\sim h^2_{10105_H}\left({M_{GUT}\over M_s}\right)^{5}\sim 10^{-22}
\ee
This is much smaller that the proton decay bound that roughly requires $h^{eff}_{10\bar 5\bar 5}\lesssim
10^{-14}$ assuming
a relevant spartner mass of about 100 GeV.

The other two terms,  $10\bar 5_H\bar 5_H$, $101010\bar 5_H$, may  be generated by instantons
that do not violate the $U(1)_X $ symmetry
and the discussion here is similar to the other cases and will be presented later.

There are further superpotential terms that may be needed for the phenomenological viability of flipped
SU(5) GUT vacua.
One of terms, namely $10_H10_H 5_H$ is necessary to give a large mass to the Higgs triplet. Indeed
such a term gives $<10> \bar 3_H 3_h$
where $\bar 3_H$ is the triplet contained in $10_H$ and $3_h$ is the standard EW triplet.
Such a term has charges (-1,5,0) and is therefore allowed by $U(1)_X$ but not by the anomalous U(1)
symmetry.
Therefore the same instanton that generates the bottom Yukawa couplings will also generate this term
and we can estimate its size as
$h_{10_H10_H5_H}\sim h_{10105_H}$ and the mass mixing term  $<10_H> \bar 3_H 3_h\sim
h_{10105_H}M_{GUT} \bar 3_H 3_h$.
{}This is the only contribution to the triplets' mass, from which we conclude that $M_T\sim h_{10105_H}
M_{GUT}$.
This implies that generically the triplets will be 1-3 orders of magnitude lighter than the GUT scale, a fact
that can spell problems with proton decay.

On the other hand a see-saw mechanism can work since its needs couplings $10_H 10 \phi$ with $\phi$
a singlet if $10_H$ and 10 have the same quantum numbers.

Note that such extra $10_H+\underline{10}_H$ multiplets may not exist in a given model. In the 437
flipped SU(5) vacua we have presented in the previous section there are either precisely 3 chiral $(10)$'s
and hence
no such Higgses, or there are two such Higgs pairs.

\subsection{SU(5) from $U(5)\times U(1)$ orientifolds}

In this case the relevant U(1) generators are
\be
Y=W\sp Q_{B-L}={Q_X+4W\over 5}\sp Q_B={Q_5-2W\over 5}\sp Q_X={5Q_1+Q_5\over 2}
\ee
Compared to flipped SU(5) the spectrum is the same, it is just the hypercharge and the identification of
particles that changed.
Here $Q_X$ does not participate in Y. In the relevant vacua we found ( and mentioned in section
\ref{mod})
$U(1)_{B-L}$ is massive. This implies that the associated gauge
symmetry is violated. We then expect a violation of the
global $U(1)_{B-L}$ symmetry from instanton effects. Typically this will break B-L to a discrete subgroup
and this
remnant subgroup, if non-trivial,  could play the role of a low energy R-symmetry.

In particular concerning the terms
$\bar 5 5_H$, $10\bar 5\bar 5$, $10\bar 5_H\bar 5_H$ that are unwanted, there are two possibilities:

(a) They are not generated by the instantons of the relevant vacuum, although the $Q_X$ symmetry is
violated.
This is equivalent to the statement that the leftover discrete symmetry forbids such terms.

(b) They are generated by instantons. Then only if their coefficients are very suppressed can the vacuum
pass
the baryon and lepton number violation constraints. This is, in principle at least, possible.

Similar statements apply to the two terms $1010 5_H$, $101010\bar 5$.
The first term  we want to be non-zero as it gives mass to top quarks. It should be generated by instantons
and we will find that it does so in some of our vacua. Then,  we will also argue that the second term
 is also generated by the same instantons and contributes non-trivially to proton decay.

We will finally analyze the potential impact of the $101010\bar 5_H$ term. If this term is generated at all, it is
generated by an
instanton and therefore its size is dependent on the instanton factor.
It contains the MSSM superpotential terms, $QQQH$ and $Q\bar UE_c H$ as well as couplings to the
heavy triplet Higgs $T$,
$T\bar U\bar UE_c$ and $T\bar UQQ$.
The triplet related term will generate after integrating out $T$ 6-fermion terms that are too suppressed to
worry us about proton decay.
The MSSM terms $QQQH$ and $Q\bar UE_c H$ give rise to dimension 4 terms proportional to ${\langle H
\rangle\over M_{GUT}}$ and are therefore innocuous.
The dimension 5 terms involve the Higgs and therefore give a suppressed contribution to proton decay as
they must go through a Higgs one-loop
 to generate the appropriate operators.

\subsection{SU(5) from $U(5)\times O(1)$ orientifolds}\label{UO}

In such vacua, as we described in section \ref{mod}, the U(1) brane is replaced with an O(1) brane,
that we will label as $O(1)_1$.
Typically, candidate Higgs pairs end in the hidden sector in such models.
This is the case for the solutions presented in section \ref{mod} where the hidden sector group is another
O(1) that we label $O(1)_2$.
Although the hidden sector groups associated to the Higgs branes can be different, our
 arguments below apply with trivial modifications. For this reason we assume an O(1) hidden
sector brane in the sequel.
 Note that in  table
\ref{ADKSmodOne}, there is another candidate Higgs pair from strings ending on the $O(1)_1$,
which may also be  interpreted as a mirror pair of $(\bar 5)$'s.
We will assume that this pair will  eventually become supermassive as it would be problematic otherwise,
and since
there are no quantum numbers available that would distinguish one the Higgses from a $(\bar 5)$.

The role of ``O(1)" groups in  selection rules requires some discussion.
First of all, there is no disk diagram that contains an odd number
of open string fields ending on any given brane. This implies that there is  a
perturbative $Z_2$ symmetry associated with any O(1) matter brane. This  symmetry  can be
viewed as the special case $N=1$ for $O(N)$ matter branes, and hence it is natural to call it
O(1), as if it were a gauge symmetry.

Non-perturbatively these $Z_2$ symmetries may be broken. Instantons may generate
zero-mode interactions $\epsilon_{i_1,\ldots i_N} \psi^{i_1}\ldots \psi^{i_N} $, which for odd  $N$
(the case of interest here; for even $N$ the $Z_2$ symmetry  acts on an odd number of
fermions as an $O(N)$ reflection, and is also broken by instantons)
clearly violate this  symmetry. Such an $\epsilon$ tensor cannot be generated perturbatively. Therefore
in general one expects  this symmetry to be broken from $O(N)$ to $SO(N)$.  This is completely
analogous to the breaking of the $O(32)$ symmetry of the ten-dimensional type-I string \cite{sen}.

The extrapolation of the $\epsilon$ tensor to $N=1$ may seem tricky, but is  best understood
by observing that for any $N$, the full contraction of the $\epsilon$ tensor is equal to $N!$, and
$1!=1$. Hence the analog of the $\epsilon$ tensor for $O(1)$ is $1$.

Alternatively, one may simply observe in examples
that  instantons exist which intersect certain $O(1)$ matter
branes an odd number of times, and hence, unlike disk diagrams,  there is no obvious
obstruction for generating the required $O(1)$-violating couplings.

The foregoing discussion might be confusing  because it mentions zero-mode
interactions involving an odd number of fermions.    However, it is only the number
of zero-modes involving  a given $O(1)$ matter brane that is odd. The {\it total}  number
of zero-modes  for a given instanton must in fact be even.  For O1 instanton
branes\footnote{To avoid confusion of instanton branes and matter branes, we use the notation ``O1"   for
the
instanton brane, and ``O(1)" for a matter brane}
(the only ones without  superfluous universal zero-modes)
this is in fact guaranteed by the K-theory constraints. These instanton branes are Euclideanized
symplectic matter branes.  The K-theory constraints imply as a necessary condition that all
symplectic matter branes must have an even number of intersections with any consistent brane
configuration. This must be true even if the symplectic brane does not itself participate
in that configuration, {\it i.e} even if it has vanishing Chan-Paton multiplicity. This  is known
as the ``probe-brane" constraint \cite{Uranga:2000xp}. Since this condition was checked for all
configurations in the database of \cite{adks}, we  cannot encounter any
O1 instanton branes with an odd number of fermions. Indeed, the additional instanton
required (in comparison to $U(5)\times U(1)$ models) to generate the down quark Yukawa couplings
must violate both $O(1)_1$  and $O(1)_2$. Note that the K-theory constraints do not impose
restrictions on O-type {\it matter} branes, so the required instantons are in principle allowed to exist.

We describe the couplings of interest below
\begin{itemize}

\item The $\bar 5 5_H$ term has charge $(0,1,1)$ under ($U(1)_5$,$O(1)_1$,$O(1)_2$) and is therefore
still perturbatively forbidden.
There can be in principle instantons that violate the O(1)'s and generate this term though.

\item The fate of $1~\bar 5 5_H$ term depends on what plays the role of the right-handed neutrino singlet.
There are three possibilities for the singlet seen in table \ref{ADKSmodOne}, namely (S,0), (V,V) and (0,S)
under ($O(1)_1$,$O(1)_2$).
It is only the choice (V,V) that makes this coupling, and therefore the right-handed neutrino mass
perturbatively allowed.

\item The $10\bar 5\bar 5$ term has charge (0,0) under   ($O(1)_1$,$O(1)_2$) and is therefore
perturbatively allowed.
This could spell a disaster as such a term strongly violates lepton and baryon number.
A vacuum in this class is viable if some other reason forbids this term. One possibility is to have three
distinct observable
O(1) branes and each $\bar 5$ family ends on a different O(1).
Because of antisymmetry the two terms $\bar U\bar D\bar D$ and $E_c LL$ must involve different families
and therefore are forbidden
perturbatively by O(1) charge conservation. However, this does not seem to be the case for the $Q\bar DL
$ term. We conclude
that in the absence of some additional discrete symmetry that forbids these terms, such vacua are ruled
out by proton decay.

\item The $10\bar 5\bar 5_H$ term is a standard Yukawa coupling that gives mass to the bottom quark
and is now perturbatively forbidden because it is charged as  $(0,1,1)$ under ($U(1)_5$,$O(1)_1$,
$O(1)_2$).
The only viable possibility is that it is generated by instantons.

\item The $10\bar 5_H\bar 5_H$  is uncharged under ($U(1)_5$,$O(1)_1$,$O(1)_2$) and is therefore
perturbatively allowed.
If we have a single higgs pair then such a term does not exist because of antisymmetry.
With more pairs then this term is non-trivial but provides mild constraints. However with more than one
Higgs pairs FCNC are
a generic problem to be addressed.

\item The $1010 5_H$ term gives masses to the top quark. It has charges $(+5,0,1)$ under ($U(1)_5$,
$O(1)_1$,$O(1)_2$)
It is perturbatively forbidden and can only be generated by instantons.

\item The $101010\bar 5_H$ term has charges $(+5,1,0)$ under ($U(1)_5$,$O(1)_1$,$O(1)_2$)
and is perturbatively forbidden.

\item Finally the $101010\bar 5$ term has charges $(+5,0,1)$ under ($U(1)_5$,$O(1)_1$,$O(1)_2$)
 the same as the previous Yukawa and later we will argue
that they are generated by the same instanton effects.

\end{itemize}

\section{Instanton zero modes}
\label{zeromodes}

As discussed previously, there are necessary Yukawa couplings that are perturbatively forbidden in orientifold $SU(5)$ models.  In Ref.~\cite{Blum}, the necessary conditions for generating these couplings were derived.  In this section, we shall briefly review these conditions and then consider other possible operators that would be induced by the same instanton.

In our consideration of the necessary conditions for instantonically inducing the forbidden Yukawa couplings, we shall concentrate on the fermionic zero-modes.  As mentioned earlier, we are only considering O1 branes as these contain no extra universal zero-modes that would need to be lifted.  As such, we should concentrate on the necessary charged-zero mode content to induce the
operator that we are interested in.  To zeroth order, the easiest way in order to determine the necessary
charged zero-mode content is merely to examine the net charge of the operator desired and then add the
minimum number of charged-zero modes to exactly compensate for the charge.   As the operator in
question,$10105_H$, has net $U(1)$ charges of (-1,+5) in terms of $(U(1)_1,
U(1)_5)$, we would expect that we need to integrate over a set of charged zero-modes with net charges (+1,-5) where the factor of five in $U(1)_5$ comes from the brane multiplicity associated with U(5).

However, we do need to keep in mind that ultimately we are evaluating disc amplitudes.  As such, we
shall examine which disc amplitudes are of interest.  As we shall be considering operators that are
perturbatively forbidden and because they should be induced directly into the superpotential, we are
interested in perturbatively allowed disc amplitudes containing  {\it exactly} two charged-zero modes.  A
quick examination of the $U(1)$ charges in table~\ref{Matter} reveals that one can write down a disc
amplitude containing the following states:  $10_{ij} \bar{\eta}^i\bar{\eta}^j$, where the $\bar{\eta}$ are
zero-modes transforming in a $\bar{5}$ of $SU(5)$ and the $SU(5)$ indices have been left explicit.  In
addition to this trilinear disc amplitude, another disc amplitude is required in order to generate the
operator containing the Yukawa coupling.  This other disc amplitude involves: $5_{Hm} \bar{\eta}^m
\nu$ where $\nu$ is a charged zero-mode stretched between the instanton brane and the
$U(1)_1$ brane.  Only considering these two classes of diagrams, we find that
\begin{equation}
S_{disc} = a~10_{ij} \bar{\eta}^i\bar{\eta}^j + b~5_{Hm} \bar{\eta}^m \nu+...
\label{naiveaction}
\end{equation}
where the coefficients $a,b$ would be determined by the explicit evaluation of the aforementioned disc
amplitudes
 and are moduli dependent and the ... refers to higher order terms.
 We have suppressed the family indices for the moment and we will return to this at the end.

 Considering the two classes of disc amplitudes only in Eq.~\ref{naiveaction}, we would find upon integration over
the set of zero-modes,
\begin{equation}
\int\prod_{i=1}^5d\bar{\eta}^i d\nu e^{-S_{disc}} \sim a^2 b\ \epsilon_{ijklm}  10_{ij} 10_{kl} 5_m,
\end{equation}
and so we would conclude that, indeed, the expectation of five charged zero-modes for the $U(5)$ stack
and
one charged zero-mode for the $U(1)$ stack is correct.  As mentioned earlier,
 this analysis was originally performed in Ref.~\cite{Blum}.

It is interesting to note that the disc amplitudes that we considered previously
 are merely the {\it lowest} order disc amplitudes possible.
In fact, there are higher order amplitudes that can be considered which involve the exact same zero-mode
content.  These higher order amplitudes would induce other higher order terms in $S_{disc}$ which, in
turn, would correspond to higher order terms induced in the superpotential.  We shall now consider the
next lowest order term induced in $S_{disc}$.

When considering the next lowest order disc amplitude the question of which fields
to consider is relatively important.  We shall restrict our attention only to the fields
 contained in table~\ref{Matter}.  The next lowest order disc amplitude involving
 $\bar{\eta} \nu$ is $10_{mn} \bar{5}^n \bar{\eta^m} \nu$.  This stems
 from the fact that $10\bar{5}$ has identical $U(1)$ charges to $5_H$ which is
 generic for these models given two assumptions.  The first assumption is that the
 Yukawa coupling contained in $10\bar{5}\bar{5}_H$ is perturbatively
 allowed.  This sets the $U(1)$ charges of $\bar{5}$ relative to $\bar{5}_H$.
 The second assumption is that the $U(1)$ charges for $5_H$ and $\bar{5}_H$ are opposite.
  This is a fairly generic phenomenon in the models that we examined.  Combining these
  two assumptions we find that the $U(1)_1$ charge of $\bar{5}$ should be the same as
   $5_H$ and the $U(1)_5$ opposite.  Therefore, one can always trade a $10 \cdot \bar{5}$ for a $5_H$ at
   the level of $U(1)$ charges.  For the other class of disc amplitude,  the one involving the charged
   zero modes $\bar{\eta}^i \bar{\eta}^j$,  the next lowest order invariant involves three additional
    fields ($5_H5_H1$) and, as such, we shall ignore it.

Including this new class of disc amplitude we find,
\begin{equation}
S_{disc} \sim a~10_{ij} \bar{\eta}^i\bar{\eta}^j + b~5_{Hm} \bar{\eta}^m \nu+ c ~ 10_{mn} \bar{5}^n
\bar{\eta^m} \nu +...
\label{a1}\end{equation}
and we find that after integrating out the charged zero-modes,

\begin{equation}
\int\prod_{i=1}^5d\bar{\eta}^i d\nu e^{-S_{disc}} \sim \epsilon_{ijklm}  \left(C_3\ 10_{ij} 10_{kl} 5_m +
C_4\ 10_{ij} 10_{kl} 10_{mn} \bar{5}^n \right),
\label{Wfin}
\end{equation}
where $C_3 = a ^ 2 b$ and $C_4 = a^2 c$.
There is no {\it a priori} reason why the coefficient will be zero, and we will proceed assuming that the
coefficient $C_4$ in Eq.~\ref{Wfin}
is non-zero and examine the potential phenomenological consequences of inducing this higher order
term.

Another important ingredient is the family structure of these terms. Assuming a single Higgs doublet, we
can obtain the family structure by the substitutions
\be
a10_{ij}\to \sum_{I=1}^3 a_s^I 10^I_{ij}\sp c10_{ij}\bar 5^i\to \sum_{I,J=1}^3c_s^{IJ}10^I_{ij}\bar 5^{J,i}
 \ee
 in (\ref{a1}) where $I,J$ are family indices, and $s$ is an index that labels different generating instanton
configurations.
 Such configurations are generated by the instanton brane wrapping different possible rigid cycles. This is
important for the structure of the
 effective couplings as it was first pointed out in \cite{bk,susybr} and subsequently discussed in \cite{Blum}.

 Taking all of this into account we can write the final results for the instanton generated couplings as
 \be
 \delta W=\sum_{s}e^{-S_s}\left[C^{IJ}_s~\ 10^I 10^J 5_H+D_s^{IJKL}~10^I 10^J 10^K \bar{5}^L\right]
 \label{trueW}
 \ee
 with
 \be
C^{IJ}_s= \sum_{IJ=1}^3 a_s^Ia_s^Jb_s\sp D_s^{IJKL}=\sum_{I,J,K,L=1}^3 a_s^Ia_s^Jc_s^{KL}
 \ee

\section{Effective Field Theory Analysis of proton decay operators}
\label{EFT}

As we have shown in Sect.~\ref{zeromodes}, the generation of perturbatively forbidden Yukawa couplings
via stringy instantons will typically generate {\it other} operators as well.  In this section, our goal is to
analyze the potential phenomenological implications of one of these additionally generated operators.
We shall consider the phenomenological implications of the operator $101010 \bar{5}$.
In particular, our goal is to compare the size of this incidentally instantonically induced
operator to the size of other operators which contribute to identical low-energy operators in an effective
field theory valid below the GUT scale.  In an effective field theory valid below the GUT scale, the operator
$101010\bar{5}$ contains two separate contributions to proton decay, namely
$QQQL$ and $UUDE$.  These are both dimension five operators that have been extensively considered
in the literature\cite{QQQL}.  We shall concentrate on $QQQL$ as the analysis for $UUDE$ is very similar.

Before we proceed with the size comparison, we first should note that $QQQL$ has some symmetry
considerations to take into account.  This operator is in the superpotential and, as such, should be
symmetric under the exchange of all indices ({\it i.e.} all $SU(3)$, $SU(2)$, and flavor indices).   If we
explicitly write the flavor indices as $Q _i Q_j Q_k L_l$ and as this term should be invariant under
$SU(3)$ and $SU(2)$ gauge transformations, the (suppressed) gauge indices corresponding to these
groups should be anti-symmetric under exchanges.  We therefore conclude that if $i=j=k$ then this term is
vanishing by gauge invariance.  This implies that $D^{iiii}$ from Eq.~\ref{trueW} is actually zero and that
$D^{1122}$ would be the leading contribution to proton decay.  We shall proceed assuming that this
coefficient is nonzero.  We also note that the symmetry considerations for $UUDE$ are different but similar
in nature.

We shall now consider other sources of $QQQL$ for effective field theories of $SU(5)$ GUT models.  In
the absence of the instantonically generated $101010\bar{5}$, the primary source
of $QQQL$ is the exchange of the triplet associated with the Higgs.  If the Yukawa's, $10
\bar{5} \bar{5}_H$ and $10105_H$, are non-zero then upon integrating
out triplet $QQQL$ is generated.  Thus, in our effective theory we have,

\be
(G_{\rm eff} + G_{\rm np}) QQQL,
\end{equation}
where $G_{\rm np}$ is the instantonically generated term and $G_{\rm eff}$ is the term arising from
integrating out the Higgs triplet.  Our goal is to compare the relative sizes of $G_{\rm eff}$ and $G_{\rm np}
$.

{}From Ref.~\cite{QQQL}, we have an estimate for the size of $G_{\rm eff}$.  The size of $G_{\rm np}$ can
be estimated using standard methods as well.  We find,
\begin{eqnarray}
G_{\rm eff} = \frac{h_u h_s}{M_T} \nonumber \\
G_{\rm np} = \frac{ \widetilde{\rm det}~e^{-S}}{M_{s}}
\end{eqnarray}
where, $h_u, h_s$ are the Yukawa couplings for the up and strange quark respectively, $M_T$ is the
mass of the triplet, and $\widetilde{\rm det}~ e^{-S}$ is an estimate of the size of the instanton contribution
that is generating the  $101010\bar{5}$, where $e^{-S}$ is the classical instanton
factor and $\widetilde{\rm det}$ stands for the determinant of fluctuations around the instanton.

As was shown earlier, the instantonic zero-modes that generate the Yukawa coupling are the same as
those that generate $101010\bar{5}$.  We therefore expect that the details of the
instantonic contribution should mostly drop out.  In the case of flipped $SU(5)$ we find,
\begin{equation}
\frac{G_{\rm np}}{G_{\rm eff}} = \frac{\widetilde{\rm det}}{\rm det} \frac{M_T}{M_{s}} \frac{1}{h_u} \sim
10^5 \frac{M_T}{M_{s}}
\label{b1}\end{equation}
where we have assumed that the ratio of determinants for the instantonic contributions will amount to only
${\cal O}(1)$ effects.

If the triplet obtains its mass from the standard mechanism described in section \ref{flsu}, then its related to
$M_{GUT}$ by the square of a
Yukawa coupling. This implies that it is several orders of magnitude below the GUT scale and therefore
the primary source of proton decay, is deadly.
If on the other hand there is another source of mass for the triplet so that it is $\gtrsim M_{GUT}$, then
(\ref{b1}) implies that the
contribution of the $101010\bar 5$ operator to proton decay is deadly. There seems to be no way out
except some form of fine tuning.

There could be several ways that such a fine tuning could arise:
\begin{itemize}

\item  The associated determinants relevant for the two coupling are hierarchically different in size, by
carefully tuning relevant moduli.

\item The generalized volumes of the relevant  instanton cycles are not very large so multi-instanton
corrections are comparable.
This may have as an effect that the effective instanton effect is much smaller than what indicated by the
one-instanton result.

\end{itemize}

 The corresponding ratio (\ref{b1}) for standard Georgi-Glashow $SU(5)$ is smaller by  a
factor of $\frac{h_u}{h_s}$ and is thus, better by a factor of $\sim 30$.
In this case of course fine-tuning is needed to make $M_T\gtrsim M_{GUT}$. In view of this the
$101010\bar 5$ operator is still highly problematic and additional fine-tuning is needed of the type described above for flipped SU(5).

Thus, if the GUT scale and the string scale are not separated by five orders of magnitude in energy, we
conclude
that these non-perturbative effects could be quite important, and can easily rule out SU(5) vacua.

\section{Search for Instanton branes in string vacua}\label{InstantonSearch}

\TABLE{
\begin{tabular}{|c||c|c|c|c|c|c|}
\hline
\hline
\multicolumn{2}{|c|}{Model Type} &  \multicolumn{3}{|c|}{All instanton branes} & \multicolumn{2}{|c|}
{Yukawa generators} \\
\hline
Nr. & Total  & $U$ & $S$ & $O$ & Correct zero modes & Solutions \\
\hline
2753 & 1136 & $4.9 \times 10^5$ & $1.5 \times 10^5$ & $4.8 \times 10^4$ & 84 & 6  \\
2880 & 1049 & $2.1 \times 10^5$ & $5.5 \times 10^4$ & $4.5 \times 10^4$ & 30 & 0  \\
6580 & 146 & $7.0 \times 10^4$ & $9680$ & $8092$ & 73 & 0  \\
14861 & 12 & $1190$ & $504$ & $0$ & 0 & 0  \\ \hline
617 & 16845 & $3.5 \times 10^6$ & $1.1 \times 10^6$ & $6.1 \times 10^5$ & 12889 & 0  \\
\hline
\hline
\end{tabular}
\label{InstantonSummary}
\caption{Summary of instanton branes}
}

In table (\ref{InstantonSummary})  we list the number of candidate instanton branes, divided into
unitary, symplectic and orthogonal, for all models combined. Only instantons of type $O$ have a chance
of having exactly the right number of zero-modes, but to get an idea of how common these are we have
listed
the other types as well. The fifth column indicates how many of all these candidates have {\it exactly} the
correct
number of zero-modes.

The first four rows in the table refer to the various kinds of $U(5) \times U(1)$ models discussed
earlier.
Here zero-modes from intersections of the instanton brane with $U(5)\times U(1)$ as well as
self-intersections were taken into account. The final step is  to find a hidden sector that cancel all tadpoles,
and does
not intersect the candidate instanton brane, so that no additional zero-modes are introduced.

This  turned out to be possible in precisely six cases, although only at a price: we had to allow chiral
hidden-observable matter.
In \cite{adks} such matter was always required to be  non-chiral, but it turns out that none of the tadpole
solutions described
admit an additional instanton brane. This is not surprising as intuition from the constructions and earlier
searches \cite{isu}
that such instanton branes are very rare in RCFT models with a high degree of symmetry as here.
By allowing chiral hidden-observable matter we enlarge the set of available models, and
hence the chance of success.

The six cases are all very similar, but not all identical, and occur for  the same MIPF as the six solutions
(without instantons)
for model 2753 described above. They have an hidden sector group with even more factors,
$O(4) \times O(3) \times O(2)^3 \times O(1)^2 \times U(1)^2$, and rather amazingly none of these
intersects the instanton brane.

\TABLE{
\begin{tabular}{|c||c|c|c|c|}
\hline
\hline
& \multicolumn{2}{c|}{Visible Sector}& \multicolumn{2}{|c|}{ }\\
\hline
Multiplicity & $U(5)$ & $U(1)$ & O1 Instanton & Hidden\\
\hline
5(4+1) & A & 0 & 0 & 0\\
3(3+0) & V$^*$ & V$^*$ & 0 & 0  \\
3(3+0) & 0 & S  & 0 & 0 \\
2(1+1) & V & V$^*$& 0 & 0  \\ \hline
1(1+0) & 0 & V & V & 0 \\
1(1+0) & V$^*$ & 0 & V & 0 \\ \hline
6 & V & 0 &  0 & V \\
6 & V$^*$ & 0 &  0 & V \\
5 & 0 & V &  0 & V \\
5 & 0 & V$^*$  &  0 & V \\
6 & Adj & 0 & 0 & 0 \\
2 & 0 & Adj & 0 & 0 \\
\hline
\hline
\end{tabular}
\label{ADKSInstanton}
\caption{The spectrum of the  model with exactly the correct instanton brane.}
}

The spectrum of this model is shown in (\ref{ADKSInstanton}), without details of the hidden sector,
and without purely hidden matter (matter with trivial $U(5)\times U(1)$ quantum numbers).
The detailed hidden sector and the observable-hidden matter is presented in appendix \ref{a}.
The
bi-fundamentals in lines $7\ldots 10$ are the chiral observable-hidden matter multiplets.
Although  their net chirality in $U(5)$ and $U(1)$ is -- necessarily -- zero, they are chiral because
they end on distinct hidden sector branes. Only after a breakdown of most of the hidden sector gauge
group
can these particles acquire a  mass.

We have searched the same models for instantons that may generate the unwanted couplings ({\it i.e.} those that violate
R-parity)
mentioned in section
\ref{yuk}, and we found none. This is not terribly surprising: ``good" instantons with precisely the correct zero modes are
very rare, and hence one may expect exact ``bad" instantons to be rare as well. In this particular case the large number of hidden
sectors is very likely to yield superfluous zero-modes, but it was not even necessary to check that, because already the
number of zero-modes from intersections with the $U(5)$ and $U(1)$ was too large. These statements are true in the exact RCFT
point in moduli space, where we do our computations. Outside that point some of the zero-modes may be lifted, but it is
possible that a kind of R-parity survives in the form of a  restriction on instanton zero modes.

The exponential suppression of the instanton contribution is determined by the size of ${1\over g^2}$, where  $g$ is the
gauge coupling. This quantity in its turn is determined by the coupling of the dilaton to the instanton brane. Since the instanton
brane is not a matter brane, there is at least a chance that the gauge coupling is large, and hence the instanton contribution
is not too suppressed.  In this particular example the ratio of the $U(5)$ and instanton brane dilaton coupling is 4.38. This means that
the instanton contribution is indeed considerably larger than those of standard model instantons (at the GUT scale), but still far too small
to give the right top quark Yukawa coupling (which should be of order 1). But as above, this statement is valid in the exact RCFT point.
In this context, such considerations are qualitative only. In order to get quantitative agreement, one would have to move far away
from the RCFT point into a region that is non-perturbative in the instanton brane coupling.


The last line in table (\ref{InstantonSummary}) describes the results for $U(5) \times O(1)$ models,
with the $O(1)$ factor treated analogously  as the $U(1)$ factor in the other models. In other
words, the column ``Correct zero modes" list instantons that would generate top quark Yukawa
couplings  if the Higgs comes from $U(5)$ and $O(1)$ intersections, just as the  $(\bar 5)$. As explained
earlier, this is an undesirable option, but the only one we can investigate without knowing the
hidden sector.

A better option would be to have an additional ``Higgs brane"  like the $O(1)_2$  factor
mentioned in section (\ref{UO}).
Note that this $O(1)_2$ is not part of the Standard Model brane configuration according to the criteria
used in \cite{adks}. These criteria only take into account chiral standard model matter (quarks and
leptons), and not the vector-like (M)SSM Higgs pair.  The configurations considered in \cite{adks}
have either two, three or four brane stacks, and include only those branes contributing to  chiral
matter. Indeed, the group $O(1)_2$ described above  came out coincidentally as a hidden sector.

The reason for organizing the search in that manner was that in general it is
undesirable to  have a separate Higgs brane (even though in this particular case it may still be  the best
option).
A separate Higgs brane  would imply that {\it all} couplings with a single
Higgs fields (and hence all Yukawa couplings) are perturbatively forbidden, and can at best be generated
non-perturbatively.  This is precisely the problem we are facing here.

Since the $O(1)_2$ brane is, by the definition of \cite{adks}, not part of the standard model
brane configuration, we do not have a systematic database at our disposal for such model. However,
as explained in section (\ref{mod}), we did perform  a complete hidden sector search for all 16845 models
in this
class.  There are a  few more case with just a single $O(1)_2$ hidden sector, but all of these
emerge from the same bulk invariant, and are closely related. All $O1$ instantons in
these examples have turned out to have an even number of zero-modes with $O(1)_2$, and
hence  cannot  generate any of the two required Yukawa couplings
(there were cases with an odd number of
$O(1)_1$ zero-modes, so the absence of them for $O(1)_2$ is accidental, and not due
to some overlooked selection rule).

Although the scan of the 16845 $U(5) \times O(1)$ models  was for just one sample of the
hidden sector per configuration, we are certain that all single-brane hidden sector were found,
since they appear first.
In other cases one might consider to use one of the various hidden sector branes as the Higgs brane.
However, this would require a  systematic enumeration of all possible hidden sectors for
each of the 16845 standard model configurations. In addition, the chances for finding
perfect solutions seem small: Not only would one have to find  two instantons, both for up and for down
type couplings,
but also  their intersections with all the other hidden sector branes would have to vanish.
With a large enough sample, solutions will probably exist, but given the success
 rate in other cases it is unlikely that the set of 16845
models from \cite{adks} is large  enough.
For these reasons we did not pursue these models further.

\section{Conclusions}

We have analyzed orientifold vacua with SU(5) gauge group, realizing SU(5) or flipped SU(5) grand
unification.
Many tadpole solution have been constructed from Gepner model building blocks using the algorithm
developed in \cite{adks}.
We found all such top-down constructions as well as tadpole-free vacua, with one extra observable brane
of the U(1) or O(1) type.
This is one small subset (but the simplest) of the SU(5) configurations found in \cite{adks}.

We gave a general analysis of possible terms in the superpotential of such vacua, up to quartic order, and
classified them according to their
fatality ( baryon and lepton violating interactions which are relevant or marginal), and usefulness
(Yukawa coupling).
We have classified which terms can or must be generated by instanton effects.
As is well known the top Yukawa's in SU(5) and the bottom in flipped SU(5) must be generated from
instantons (in the absence of fluxes).

In flipped SU(5) vacua, B-L cannot be anomalous as it participates in
the hypercharge.
It forbids all dangerous terms, but it is necessarily broken when the SU(5) gets broken at the GUT scale.
We have estimated that the proton decay generated is typically small.

In U(5)$\times$U(1)  vacua, instanton effects must generate the top Yukawa couplings, and at the same
time they break the B-L symmetry.
Successful vacua, have either a $Z_2$ remnant of the B-L symmetry acting as as R-parity and forbidding
the dangerous terms, or
such term may have exponentially suppressed instanton contributions. In the second case they are viable
if the exponential factors are
sufficiently suppressed.  We provide several tadpole solutions of the first
case where instantons generate the top Yukawa's, but preserve a $Z_2$ R-symmetry.

Finally U(5)$\times O(1)$ vacua are problematic on several grounds and need
extra symmetries beyond those that are automatic, in order to have a chance of not
being outright excluded. This is related to the absence of natural R-symmetries or gauge symmetries that
 will forbid the dangerous low-dimension baryon-violating interactions.

A generic feature of all SU(5) vacua is that the same instanton that generates the non-perturbative quark
Yukawa coupling also generates
the $10~10~10~\bar 5$ in the superpotential. This is a second source of proton decay, beyond the
classic one emanating from the
Higgs triplet times the appropriate Yukawa coupling. Generically, the size of this contribution to proton
decay
is $10^5 ~{M_T\over M_s}$ larger than the conventional source in the flipped SU(5) model, ($M_T$ is the
triplet Higgs mass).
This  signals severe phenomenological trouble
and calls for important fine-tuning. In the SU(5) case, the size is 30 times smaller, but that does not evade
the need for fine-tuning.

 We have searched for appropriate instantons that would generate the perturbatively forbidden quark
 Yukawa couplings in the SU(5) vacua we have constructed. We found the appropriate instantons with the
correct number of zero modes in 6 relatives
 of the spectrum Nr. 2753. We have also searched for all other instantons that could generate the bad
terms in the superpotential and found none.
 This translates into the existence of a $Z_2$ R-symmetry that protects from low-dimension baryon and
lepton-violating couplings.

A related formalism that provides orientifold vacua with a non-perturbative description for some of their
features is F-theory, \cite{f}.
This is a new area for model building and recently bottom-up constructions of SM stacks of D7 branes
were explored, \cite{f-recent,f1}. Global constructions are in their infancy, \cite{gf1,gf2} but despite this,
phenomenologically interesting global GUT
vacua were recently described in \cite{gf2}.

Like  orientifolds as long as the appropriate U(1) symmetry that forbids the top Yukawa coupling
is present, then the coupling can be generated only by instantons. In such a case, our discussion,
estimates and conclusion remain unaltered.
However, in F-theory there is the option of breaking the offending U(1) symmetry
non-perturbatively by considering enhanced symmetry singularities. This now allows the top Yukawa coupling at triple intersections
of appropriate singularities. The offending $101010\bar 5$ term may be now generated via three possible sources:
(a) Mediation by higher triplets (for example KK triplets). (b) Potential $D_3$-instantons effects.
(c) String instantons stretched
between four appropriate divisors.

The  (a) contribution is phenomenologically dangerous  and an idea to avoid it has been advanced in
\cite{f-recent} but putting  the up and down Higgses on different divisors.

Contribution (b) is no-longer  guaranteed to  exist but if it does, it is no longer related to the  top Yukawas.
 It is generically exponentially suppressed. Our arguments in   section \ref{EFT} imply that if the coupling generated here is much smaller
 than about $10^{-12}$ in string units then we do not need to worry about it. Otherwise a detailed analysis is necessary.

Contribution (c) is also generically exponentially suppressed. The reason is that four-point intersections of divisors are non-generic.
However, the quantitative statements and constraints in  (b) are also valid in this case.

Finally we should mention that in special cases extra PQ-like (anomalous) symmetries may forbid the
 $101010\bar 5$ term, while allowing the top Yukawa.
An example based on such a symmetry emanating from $E_6$ was described in \cite{f-recent}.


\vspace{1.2 in}
\addcontentsline{toc}{section}{Acknowledgments}

\noindent {\Large\bf Acknowledgements} \newline

We are grateful to A. Arvanitaki, R. Blumenhagen, S. Dimopoulos, W. Lerche, J. March-Russell, S. Raby, J. Rizos, K.
Tamvakis, A. Uranga, C. Vafa, J. Wells, T. Weigand, A. Wingerter and F. Zwirner for
valuable discussions and correspondence.

 This work was  partially supported
 by  a European Union grant FP7-REGPOT-2008-1-CreteHEPCosmo-228644,
  and a CNRS PICS grant \# 4172.
 The
work of A.N.S. has been performed as part of the program
FP 57 of the Foundation for Fundamental Research of Matter (FOM),
and has been partially
supported by funding of the Spanish Ministerio de Ciencia e Innovaci\'on, Research Project
FPA2008-02968.

\vskip 2cm
\appendix
\renewcommand{\theequation}{\thesection.\arabic{equation}}
\section{APPENDIX\label{a}}

In this appendix we present the exact spectrum of one of the models that have an exact instanton
brane (the other models are nearly identical).  The first column gives an ad-hoc number we use for
referring to the various massless states.
The second column gives the total multiplicity
(for representation plus its conjugate), the last column the
chiral multiplicity ({\it i.e} the multiplicity for the representation minus its conjugate).

The spectrum  is divided in the table into the following segments:
Quarks and leptons (1-3), the third row contributes 5 symmetric tensors of $U(1)$, with a net chirality 3;
these can play the
r\^ole of right-handed singlet neutrinos),  the Higgs pair (4), the instanton zero modes (5-6), chiral
observable-hidden matter (7-12),
 non-chiral
observable-hidden matter (13-16),  non-chiral observable rank two tensors (17-20), chiral matter within
the hidden sector (21-29), and non-chiral matter
within the hidden sector (30-49).

\normalsize

The chiral exotics may acquire masses via symmetry breaking in the hidden sector.
 In view of the size of the hidden sector such a analysis lies beyond the scope of the present paper.

\scriptsize
\TABLE{
\begin{tabular}{|c|c||c|c|c|c|c|c|c|c|c|c|c|c||c|}
\hline
\hline
Num.&Mult. & $U(5)$ & $U(1)$ & $O(1)_1$ & $U(1)_1$ & $O(3)_1$ & $U(1)_2$ & $O(2)_1$ & $O(4)$ &
$O(2)_2$ & $O(2)_3$ & $O(1)_2$ & O1 & Chir.\\
 &        &               &               &               &              &               &               &              &    &   &    &   &  Inst. & \\
\hline
1&     5 &  A & 0 & 0 & 0 & 0 & 0 & 0 & 0 & 0 & 0 & 0 & 0 & 3\\
2&     3 &  V & V & 0 & 0 & 0 & 0 & 0 & 0 & 0 & 0 & 0 & 0 & -3\\
3&     5 &  0 & S & 0 & 0 & 0 & 0 & 0 & 0 & 0 & 0 & 0 & 0 & 3 \\ \hline
4&     2 &  V &  V$^*$ & 0 & 0 & 0 & 0 & 0 & 0 & 0 & 0 & 0 & 0 & 0\\ \hline
5&     1 &  V & 0 & 0 & 0 & 0 & 0 & 0 & 0 & 0 & 0 & 0 & V & -1\\
6&     1 &  0 & V & 0 & 0 & 0 & 0 & 0 & 0 & 0 & 0 & 0 & V & 1\\ \hline
7&     1 &  V & 0 & 0 & 0 & 0 & 0 & 0 & V & 0 & 0 & 0 & 0 & 1\\
8&     1 &  V & 0 & 0 & 0 & 0 & 0 & 0 & 0 & 0 & V & 0 & 0 & 1\\
9&     2 &  V & 0 & 0 & 0 & 0 & 0 & V & 0 & 0 & 0 & 0 & 0 & -2\\
10&     3 &  0 & V & 0 & 0 & 0 & 0 & 0 & 0 & 0 & V & 0 & 0 & -1\\
11&     1 &  0 & V & 0 & 0 & 0 & 0 & V & 0 & 0 & 0 & 0 & 0 & 1\\
12&     2 &  V & 0 & 0 & 0 & 0 & 0 & 0 & 0 & 0 & 0 & V & 0 & -2\\ \hline
13&     4 &  0 & V & 0 & 0 & 0 & 0 & 0 & 0 & 0 & 0 & V & 0 & 0\\
14&     2 &  0 & V & 0 & 0 & 0 & 0 & 0 & V & 0 & 0 & 0 & 0 & 0\\
15&     4 &  0 & V & 0 & 0 & 0 & V & 0 & 0 & 0 & 0 & 0 & 0 & 0\\
16&     2 &  0 & V & 0 & 0 & 0 & 0 & 0 & 0 & V & 0 & 0 & 0 & 0\\ \hline
17&     2 &  S & 0 & 0 & 0 & 0 & 0 & 0 & 0 & 0 & 0 & 0 & 0 & 0\\
18&     4 &  0 & A & 0 & 0 & 0 & 0 & 0 & 0 & 0 & 0 & 0 & 0 & 0\\
19&     6 &  Ad & 0 & 0 & 0 & 0 & 0 & 0 & 0 & 0 & 0 & 0 & 0 & 0\\
20&     2 &  0 & Ad & 0 & 0 & 0 & 0 & 0 & 0 & 0 & 0 & 0 & 0 & 0\\ \hline
21&     1 &  0 & 0 & 0 & 0 & 0 & S & 0 & 0 & 0 & 0 & 0 & 0 & -1\\
22&     1 &  0 & 0 & 0 & 0 & 0 & A & 0 & 0 & 0 & 0 & 0 & 0 & 1\\
23&     1 &  0 & 0 & 0 & 0 & 0 & V & V & 0 & 0 & 0 & 0 & 0 & 1\\
24&     1 &  0 & 0 & 0 & 0 & 0 & V & 0 & V & 0 & 0 & 0 & 0 & 1\\
25&     1 &  0 & 0 & 0 & V & 0 & 0 & 0 & 0 & V & 0 & 0 & 0 & -1\\
26&     1 &  0 & 0 & 0 & 0 & 0 & V & 0 & 0 & V & 0 & 0 & 0 & 1\\
27&     1 &  0 & 0 & 0 & V & 0 & 0 & 0 & 0 & 0 & V & 0 & 0 & 1\\
28&     1 &  0 & 0 & 0 & 0 & 0 & V & 0 & 0 & 0 & V & 0 & 0 & 1\\
29&     2 &  0 & 0 & 0 & 0 & 0 & V & 0 & 0 & 0 & 0 & V & 0 & -2\\ \hline
30&     2 &  0 & 0 & 0 & 0 & V & V & 0 & 0 & 0 & 0 & 0 & 0 & 0\\
31&     1 &  0 & 0 & V & 0 & V & 0 & 0 & 0 & 0 & 0 & 0 & 0 & 0\\
32&     2 &  0 & 0 & 0 & V & 0 &  V$^*$ & 0 & 0 & 0 & 0 & 0 & 0 & 0\\
33&     1 &  0 & 0 & 0 & 0 & A & 0 & 0 & 0 & 0 & 0 & 0 & 0 & 0\\
34&     1 &  0 & 0 & 0 & 0 & 0 & Ad & 0 & 0 & 0 & 0 & 0 & 0 & 0\\
35&     1 &  0 & 0 & 0 & 0 & 0 & 0 & S & 0 & 0 & 0 & 0 & 0 & 0\\
36&     1 &  0 & 0 & 0 & 0 & 0 & 0 & V & V & 0 & 0 & 0 & 0 & 0\\
37&     1 &  0 & 0 & 0 & 0 & 0 & 0 & 0 & S & 0 & 0 & 0 & 0 & 0\\
38&     2 &  0 & 0 & V & 0 & 0 & 0 & V & 0 & 0 & 0 & 0 & 0 & 0\\
39&     2 &  0 & 0 & 0 & 0 & V & 0 & V & 0 & 0 & 0 & 0 & 0 & 0\\
40&     2 &  0 & 0 & 0 & 0 & V & 0 & 0 & V & 0 & 0 & 0 & 0 & 0\\
41&     1 &  0 & 0 & V & 0 & 0 & 0 & 0 & 0 & V & 0 & 0 & 0 & 0\\
42&     1 &  0 & 0 & 0 & 0 & 0 & 0 & 0 & 0 & S & 0 & 0 & 0 & 0\\
43&     1 &  0 & 0 & 0 & 0 & 0 & 0 & 0 & V & V & 0 & 0 & 0 & 0\\
44&     1 &  0 & 0 & 0 & 0 & V & 0 & 0 & 0 & 0 & 0 & V & 0 & 0\\
45&     1 &  0 & 0 & 0 & 0 & 0 & 0 & 0 & 0 & 0 & A & 0 & 0 & 0\\
46&     2 &  0 & 0 & 0 & 0 & 0 & 0 & 0 & 0 & 0 & V & V & 0 & 0\\
47&     2 &  0 & 0 & 0 & 0 & 0 & 0 & 0 & 0 & 0 & 0 & A & 0 & 0\\
48&     2 &  0 & 0 & 0 & 0 & 0 & 0 & 0 & V & 0 & 0 & V & 0 & 0\\
49&     1 &  0 & 0 & 0 & 0 & 0 & 0 & 0 & 0 & V & 0 & V & 0 & 0\\
\hline
\hline
\end{tabular}
\label{ADKSInstantonFull}
\caption{The complete spectrum of the  model with exactly the correct instanton brane.}}

\end{document}